\documentclass[11pt,a4paper]{article}
\usepackage{jcappub}
\usepackage[utf8]{inputenc}
\usepackage[a4paper]{geometry}
\usepackage{units}
\usepackage{amsmath}
\usepackage{graphicx}
\usepackage{todonotes}
\usepackage{booktabs}
\usepackage{multirow}
\usepackage{hyperref}

\title{Combined detection of supernova neutrino signals}
\author[a]{A. Sheshukov}
\author[a,b]{A. Vishneva}
\author[c]{A. Habig}
\affiliation[a]{Joint Institute for Nuclear Research\\
 Dubna, Moscow region, 141980, Russia
}
\affiliation[b]{St. Petersburg Nuclear Physics Institute NRC Kurchatov Institute\\ Gatchina, 188350, Russia}
\affiliation[c]{Department of Physics and Astronomy, University of Minnesota Duluth\\ Duluth, Minnesota, 55812, USA}
\emailAdd{andrey.sheshukov@jinr.ru}
\arxivnumber{2107.13172}

\begin{document}

\abstract{
    Supernova neutrino detection in neutrino and dark matter experiments is usually implemented as a real-time trigger system based on counting neutrino interactions within a moving time window. The sensitivity reach of such experiments can be improved by taking into account the time profile of the expected signal. We propose a shape analysis of the incoming experimental data based on a log likelihood ratio variable containing the assumed signal shape. This approach also allows a combination of potential supernova signals in different detectors for a further sensitivity boost. The method is tested on the NOvA detectors to study their combined sensitivity to the core-collapse supernova signal, and also on KamLAND, Borexino and SK-Gd as potential detectors of presupernova neutrinos. Using the shape analysis enhances the signal significance for supernova detection and prediction, as well as the sensitivity reach of the experiment. It also extends the supernova prediction time when applied to the presupernova neutrino signal detection. Enhancements achieved with the shape analysis persist even in the case when the actual signal doesn't match the expected signal model.

}

\maketitle
%\newpage
%\tableofcontents

\newpage
\section{Introduction}

A core-collapse supernova progenitor star produces a flux of neutrinos that gradually increases  during the last weeks prior to the explosion (presupernova signal), then a short and intense neutrino burst during the core collapse process.

The energy spectrum, flavor composition, and time profile of this neutrino signal depend on the physics processes in the supernova as well as the neutrino properties, providing a unique opportunity to probe models and study effects that would otherwise be experimentally inaccessible~\cite{scholberg_supernova_2012}.
On the other hand, the expected (pre)supernova neutrino signals have common features  across various models, allowing a robust supernova burst indication in real time to provide an early warning for the electromagnetic supernova explosion. 

Many neutrino and dark matter detectors are potentially sensitive to neutrino signals from supernovae. Supernova detection systems in such experiments are designed to identify the presence of the supernova signal based on detected neutrino interaction candidates in one or many detectors. 

The detection of the supernova signal would benefit from the coordinated effort of existing neutrino experiments. 
Such coordination started historically as a post-factum comparison of the data between several experiments after the detection of SN1987A~\cite{SN1987Areview1989}, was later implemented as a real-time coincidence server of the SuperNova Early Warning System (SNEWS)~\cite{snews1}, and now is being designed as a new generation global system (SNEWS2.0)~\cite{snews2021} featuring more complex combinations of the measured signals.

%There exist several approaches \cite{Casentini_2018, lamoureux_identification_2021} for the identification of the supernova signal searching for the short neutrino burst, however these methods are only applicable for the core-collapse signal. Moreover, they are designed only for the low-background cases, so they cannot be applied in different background conditions. \todo{Rephrase better. They are not physically motivated, just use some functions}

In this work we propose a method for identification and calculation of the significance of supernova occurrence based on the knowledge of the expected signal shape and background rate versus time. 
This \textit{shape analysis}~(SA) method was first applied in the supernova triggering system of the NOvA experiment~\cite{nova_collaboration_supernova_2020}.
However, the method is general enough to be used in the searches for both the short neutrino signal from the core-collapse and the long presupernova signal. It is also applicable to calculate the combined significance of different experiments and/or detection modes.

The shape analysis method is described in section~\ref{seq:method} in comparison with the commonly used \textit{counting analysis}~(CA), which takes into account only the total number of detected neutrinos within a given time window. Section~\ref{seq:nova} describes the application of the method to the detection of core-collapse neutrino burst using the example of the NOvA detectors. This illustrates both the advantages of this method in comparison to the counting analysis and the benefit of the combined analysis of separate detectors. 
In section~\ref{seq:presn} we apply the method to several experiments (KamLAND, Borexino, SK-Gd) sensitive to the pre-supernova neutrino signal. These detectors were chosen to illustrate the method because their presupernova sensitivity is comparable, and thus the benefits of the joint analysis are more visible. It also demonstrates a combination of the two detection channels within one detector: neutron events and delayed coincidences in SK-Gd detector.

\newpage

\newcommand{\F}{\mathcal{F}}
\newcommand{\FT} [1]{\F\left\{#1\right\}}
\newcommand{\FTi}[1]{\F^{-1}\left\{#1\right\}}
\section{Shape analysis method}
\label{seq:method}
\subsection{Hypothesis test}

Supernova neutrino detection is usually implemented as a real-time trigger system to provide an immediate reaction to the occurrence of signal-like structures in a stream of data. This real-time detection regime implies continuous computation of the probability that the observed data at some point in time is consistent with the expected neutrino signal coming from the late stages of the evolution of massive stars: either the burst of neutrinos from a core collapse, or the cooling of a nearby red supergiant core in the Si-burning presupernova phase.  
%This statistical analysis can be done in a way that allows the combination of significances from multiple detectors.  

The most thorough approach for locating the signal in the presence of background would be fitting the measured data with the supernova signal parameters (distance, mass, starting time etc.). 
However, this approach has limited efficacity because existing simulations might not cover all the physics possibilities and hypotheses. 
Moreover, performing fits to multiple models in a large parameter space is computationally intensive and appears not feasible for the case of the real-time supernova detection, as such evaluation has to be done repeatedly for each new portion of data received. 
Therefore we consider a more simple and robust procedure of hypotheses test, measuring the supernova significance as the deviation from the background and avoiding the scan of all supernova signal parameters, except for the starting time $t^{*}$.

Consider an experiment that measures timestamps of individual neutrino interaction candidates $\{t_i\}$. The task of the supernova detection system is to tell if the observed data contains a signal from a supernova and to evaluate the starting time of that astrophysical event. Assuming the supernova starting time $t^\ast$, such system scans the incoming data and performs statistical tests to estimate the significance of the signal starting at $t^\ast$ in the considered data slice.
To discriminate between two hypotheses, $H_0$ (background only) and $H_1$ (background + supernova), one has to define a \emph{test statistic} as a function of the incoming data: $\ell (\{t_i\})$.
%A \emph{test statistic} is calculated as a function of the incoming data $\ell (\{t_i\})$ and used to discriminate between two hypotheses $H_0$ (background only) and $H_1$ (background + supernova).
%\subsection{Significance}
    
The significance of observing deviations from the null hypothesis $H_0$ using the test statistic $\ell$ is defined by the tail of the distribution:
\begin{equation}
    p(\ell) = \int\limits_{\ell}^{\infty} P(\ell' | H_0)\,d\ell' .
\label{eq:pvalue}
\end{equation}
For convenience this can be converted to a $z$-score, defined as 
\begin{equation}
    z(\ell) = \Phi^{-1}(1-p(\ell)) ,
\label{eq:p2z}
\end{equation}
where $\Phi$ is the cumulative standard normal distribution function.  This z-score is equivalent to that derived in Gaussian statistics as the number of sigma away from the mean.  
%\subsection{Triggering}

The supernova triggering system would react on the signal present in the data when the significance z-score exceeds a defined threshold $z_{thr}$. Thus the efficiency of the supernova detection under hypothesis $H$ is defined by the probability distribution above this threshold:
\begin{equation}
\varepsilon(H) = \int\limits_{z_{thr}}^\infty P(z|H)\; dz .
\label{eq:trigger_eff}
\end{equation}

\subsection{Choice of the test statistic function}
\subsubsection*{Counting analysis (CA)}

The most common approach for detection of supernova (and presupernova) neutrinos is to count the number of events $n$ within a time window $[t^* , t^*+\Delta t]$, where $t^*$ is the assumed supernova starting time, and use it as the test statistic:
\begin{equation}
\ell_{CA}(t^*,\Delta t, \{t_i\}) = n \equiv \sum_i w\left(t_{i}-t^*,\Delta t\right), 
\end{equation}
where $w(t, \Delta t)$ is a window function
$$
w(t, \Delta t) = \begin{cases} 
1, & t\in[0,\Delta t], \\
0, & \text{ otherwise.}
\end{cases}
$$
The advantages of this approach are simplicity of the significance calculation and independence of the expected signal model.
However, as the CA does not account for the expected time distribution of the signal events, it produces a maximal significance around the time of  signal maximum, later than the actual starting time of the signal. Moreover, analyzing only the total event number within a certain window provides a risk of misidentifying background fluctuations or detector instabilities as a signal. 

\subsubsection*{Shape analysis (SA)}

A more efficient discrimination between signal and background hypotheses can be achieved by taking into account the shape of the expected signal and background rates over time. 
While we avoid using the curve fit procedure with many signal parameters, it's possible to maximize the discrimination for a specific signal model by using the log likelihood ratio \cite{Cowan_2011} for hypotheses $H_0$ and $H_1$ as the test statistic function:

\begin{equation}
    \ell_{SA}(t^*,\{t_i\}) = \log\frac{P(\{t_i\} | H_1)}{P(\{t_i\} | H_0)} = \sum_i \log\left(1+\frac{S(t_i-t^*)}{B(t_i)}\right) ,
    \label{eq:llr_calc}
\end{equation}
where $B(t)$ is the background event rate, and $S(t)$ is the expected signal event rate over time, relative to supernova start time $t^*$.

As this approach maximizes the detection power for a specific signal with fixed supernova distance $d^*$ and other model parameters, except for $t^*$, the discrimination will be suboptimal for supernovae at distances different from $d \approx d^*$.
It is reasonable to choose $d^*$ at which the detection efficiency is close to $50\%$, so the discrimination is optimal in the transition region between zero and $100\%$ efficiency, while the sensitivity to the signal is too low at larger distances and good enough at shorter ones, regardless of the choice of $d^*$.

Note that SA approach is equivalent to CA if the expected signal shape is chosen to be a constant within a counting window: $S(t)\sim w(t,\Delta t)$.
%This approach depends on the signal model and expected signal shape, and will be referred here as the ``shape analysis'' (SA).

\subsection{Test statistic distribution}

Calculation of the significance values \eqref{eq:pvalue} and \eqref{eq:p2z} requires knowing the probability distribution of the test statistic under a given hypothesis $P(\ell | H)$. 

For a hypothesis $H$ predicting the event rate $r(t)$, the number of events $n$ within the considered time window $[t^*, t^*+\Delta t]$ is a Poisson random number around the expected rate:
\begin{equation}
    P(n|H) = \frac{e^{-R} R^n}{n!},\;\text{ where } R=\int\limits_{t^*}^{t^*+\Delta t} r(t)\,dt .
    \label{eq:pN}
\end{equation}

%Obviously, as this will be the expression for the for the CA method, where the  test statistic is $n$
%$$
%P(\ell_{CA}| H) = \mathrm{Poisson}(\ell|\lambda=R)
%$$

Test statistics in both CA and SA are additive functions of the timestamps of neutrino candidate data $\{t_i\}$: $\ell(\{t_i\}) = \sum_i \ell(t_i)$. Thus the total test statistic distribution can be expressed as a weighted sum of contributions for each $n$:
\begin{equation}
    P(\ell| H) = \sum_{n=0}^{\infty} P(\ell|n, H)\cdot P(n| H) .
    \label{eq:pl_expansion}
\end{equation}

%For both CA and SA the test statistic $\ell$ is additive function of the event timestamps $\ell(\{t_i\}) = \sum_i \ell(t_i)$. 
Assuming that all the individual events  are causally independent, we can express $P(\ell|n, H)$ as a convolution with a single event distribution $P(\ell|1,H)$:
\begin{equation}
    P(\ell|n, H) =P(\ell|n-1,H)*P(\ell|1,H) ,
\end{equation}
which, using the convolution theorem, becomes:
\begin{equation}
    P(\ell|n, H) = \FTi{\Phi^n(z)} ,
    \label{eq:plN}
\end{equation}
where $\Phi(z) = \FT{P(\ell|1, H)}$ is a characteristic function for $P(\ell|1, H)$, and $\FT{}$ and $\FTi{}$ denote direct and inverse Fourier transformations.

The distribution of the test statistic value for a single event with timestamp $t$ is defined as:

\begin{equation}
P(\ell|1, H) = \int\limits_{t^*}^{t^*+\Delta t} dt  P(t|1, H) \int\limits_0^\infty  d\ell'  \delta(\ell' - \ell(t)).
\label{eq:pl1H}
\end{equation}
and the distribution of a single timestamp within the time window is defined by the event rate, predicted by hypothesis $H$: $P(t|1, H) = r(t)/R$. These expressions are used for the numerical computation of the characteristic function $\Phi(z)$.

Putting \eqref{eq:pN} and \eqref{eq:plN} into \eqref{eq:pl_expansion} we get:
\begin{equation}
    P(\ell|H) = \sum\limits_{n=0}^\infty \frac{e^{-R}R^n}{n!}\FTi{\Phi^n(z)} = \FTi{ e^{-R} \sum\limits_{n=0}^\infty \frac{(R\Phi(z))^n}{n!}} .
\end{equation}
Finally, combining the series into to an exponent, we get the expression for the test statistic distribution:
\begin{equation}
 P(\ell|H) = \FTi{ e^{R[\Phi(z)-1]}}
 \label{eq:pell_h_fourier} .
\end{equation}

\subsection{Combining measurements}
In case of multiple experiments each using their own test statistic functions $\{\ell_n(t)\}$, their combination is nontrivial.
However, if each of $\ell$ is a log likelihood ratio $\ell(t) = \log \frac{P(t|H_1)}{P(t|H_0)}$ and all the experiments use the same hypotheses $H_0$ and $H_1$, these values are additive:
$$
\ell_{comb} = \sum\limits_{n=1}^{N_{exp}} \ell_n(\{t_i^n\}) = \sum\limits_{n=1}^{N_{exp}} \sum_i \ell_n(t_i^n) ,
$$
where $N_{exp}$ is number of experiments to combine.

In this case the distribution of combined test statistic $P(\ell_{comb}|H)$ becomes a convolution of distributions for the individual experiments, so \eqref{eq:pell_h_fourier} becomes:
\begin{equation}
    P(\ell_{comb}|H) = \FTi{\prod_{n=1}^{N_{exp}} \exp\left(R_n [\Phi_n(z)-1]\right)} .
    \label{eq:pell_comb}
\end{equation}

\subsection{Implementation}

The shape analysis method described above is implemented as a python package and is available at~\cite{sn_stat}.

Briefly, the computation of the resulting significance requires the following preparation steps:
\begin{enumerate}
    \item Define the background $B$ and expected signal $S$ rates.
    \item Compute the single event distribution $P(\ell|1, H_0)$ using \eqref{eq:pl1H}, and its image $\Phi(z)$ with the fast Fourier transformation. 
    \item Calculate the test statistic distribution for the background-only hypothesis $P(\ell|H_0)$ using eqs. \eqref{eq:pell_h_fourier} or \eqref{eq:pell_comb} in case of many experiments.
\end{enumerate}

Then the data analysis can be done, based on the events timestamps $\{t_i\}$  for each considered supernova starting time $t^*$:
\begin{enumerate}
    \item[4.] Calculate the test statistic value $\ell(t^*,\{t_i\})$ via \eqref{eq:llr_calc}
    \item[5.] Evaluate the significance corresponding to the obtained value $\ell$  using $P(\ell|H_0)$ via \eqref{eq:pvalue}, \eqref{eq:p2z}
\end{enumerate}
This results in a time series of significance values, representing the likelihood that a supernova following the $H_1$ hypothesis started at each point in time.

Computationally the shape analysis method is more complicated than the counting analysis, which relies only on the Poisson distribution. However, the calculation of the test statistic distribution needs to be done only when the background level (or expected signal) has changed, so in practice these reevaluations can be done with large time intervals, leading to a smaller computational load and latency on average. For example, in the case of the NOvA supernova triggering system~\cite{nova_collaboration_supernova_2020} the reevaluation of the background level and recalculation of the null hypothesis distribution is performed every \unit[10]{minutes}, as that is the characteristic time over which background rates in NOvA are stable.

\subsection{Example}

\begin{figure}[!htb]
    \centering
    \includegraphics[scale=0.5]{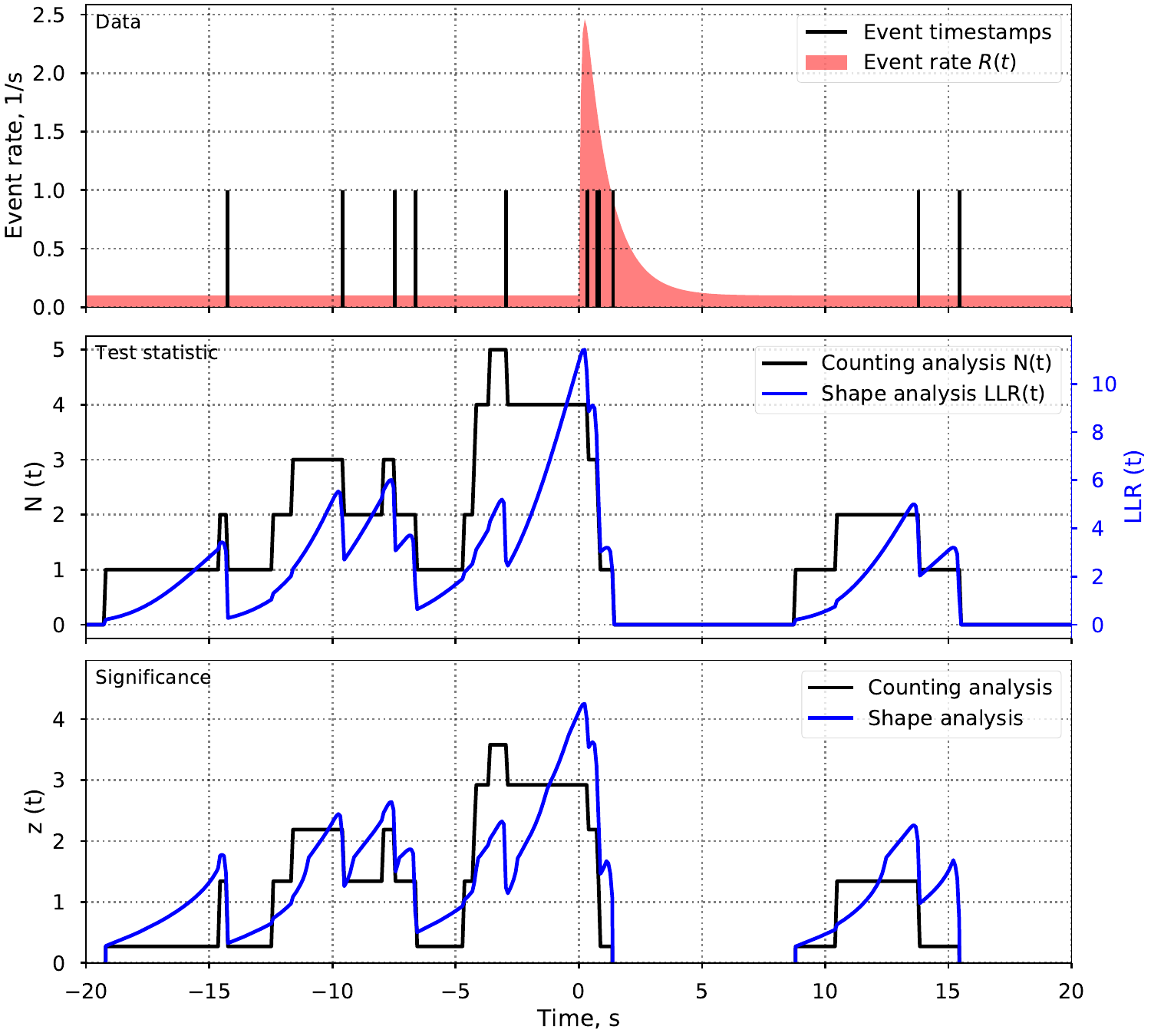}
    \caption{Example of applying the shape analysis and the counting analysis methods to a set of measured events. The top panel shows the expected event rate over time (filled area) and sampled events timestamps (vertical lines).  The middle panel plots  test statistic values vs. assumed time of the supernova for both methods: number of events N(t) for CA, and log likelihood ratio LLR(t) for SA. The bottom panel shows the calculated significance of the expected signal starting at each time.}
    \label{fig:ca_sa_example}
\end{figure}

Figure~\ref{fig:ca_sa_example} shows an example of such computation for an experiment with a background rate $B=\unit[0.1]{event/s}$ and a signal with total expected $\int S(t) dt = \unit[3]{events}$, injected at time $T=0$. For the signal shape we used an analytical expression to imitate a typical core-collapse anti-neutrino signal: 
\begin{equation}
S(t)\sim (1-e^{-t/\tau_0})\cdot e^{-t/\tau_1},    
\label{eq:simple_param}
\end{equation}

with time parameters $\tau_0 = \unit[0.1]{s}$, $\tau_1 = \unit[1]{s}$. 

The length of the time window for both analyses is set to \unit[5]{s}, so each neutrino interaction (vertical lines in the top panel) increases the test statistic value for the assumed supernova start time within \unit[5]{s} in the past. For the counting analysis this impact is constant within the whole time window, while for the shape analysis it is defined by the signal and background shapes, thus giving a more precise prediction of the start time, and a higher maximal significance.

%\subsection{Rate uncertainty}
%\todo[inline]{This is not used in this work. Maybe needs to be removed}
%In case the event rate $R$ is defined with an estimated mean value $\bar{R}$ and the uncertainty $\sigma_R$, we should integrate the \eqref{eq:pell_h_fourier} with the distribution (in Bayesian sense) of the true value $R$:
%\begin{equation*}
%    P(\ell|H) = \int dR\cdot P(\ell| R)\cdot P(R|H) = \int \frac{dR}{\sqrt{2\pi}\sigma_R}\;
%    \FTi{\exp\left(R[\Phi(z)-1]-\frac{(R-\bar{R})^2}{2\sigma_R^2}\right)} 
%\end{equation*}

%And by extracting the full square 
%$$
%R\phi - \frac{(R-\bar{R})^2}{2\sigma_R^2} =  -\frac{(R-\bar{R}-\sigma_R^2\phi)^2}{2\sigma_R^2} + \bar{R}\phi+\frac{\sigma_R^2\phi^2}{2}
%$$
%and integrating over $dR$ we get:
%\begin{equation}
%    \label{eq:pell_rdr}
%    P(\ell|H) = \FTi{\exp\left( \bar{R} [\Phi(z)-1] + \frac{\sigma_R^2}{2}[\Phi(z)-1]^2\right)}
%\end{equation}

%\begin{equation}
%    P(\ell_{comb}|H) = \FTi{\prod_{n=1}^{N_{exp}} \exp\left(\bar{R}_n [\Phi_n(z)-1] + \frac{\sigma_{R_n}^2}{2}[\Phi_n(z)-1]^2\right)}
%    \label{eq:pell_comb_rdr}
%\end{equation}

\newpage
\section{Supernova neutrino signal in NOvA detectors}
\label{seq:nova}

The shape analysis method described above has been implemented in the supernova triggering system of the NOvA experiment~\cite{nova_collaboration_supernova_2020}. Here we consider the NOvA case to illustrate the performance of this method for the core-collapse neutrino signal detection in combined analysis of multiple detectors.

\begin{figure}[htb]
    \centering
    \includegraphics[scale=0.5]{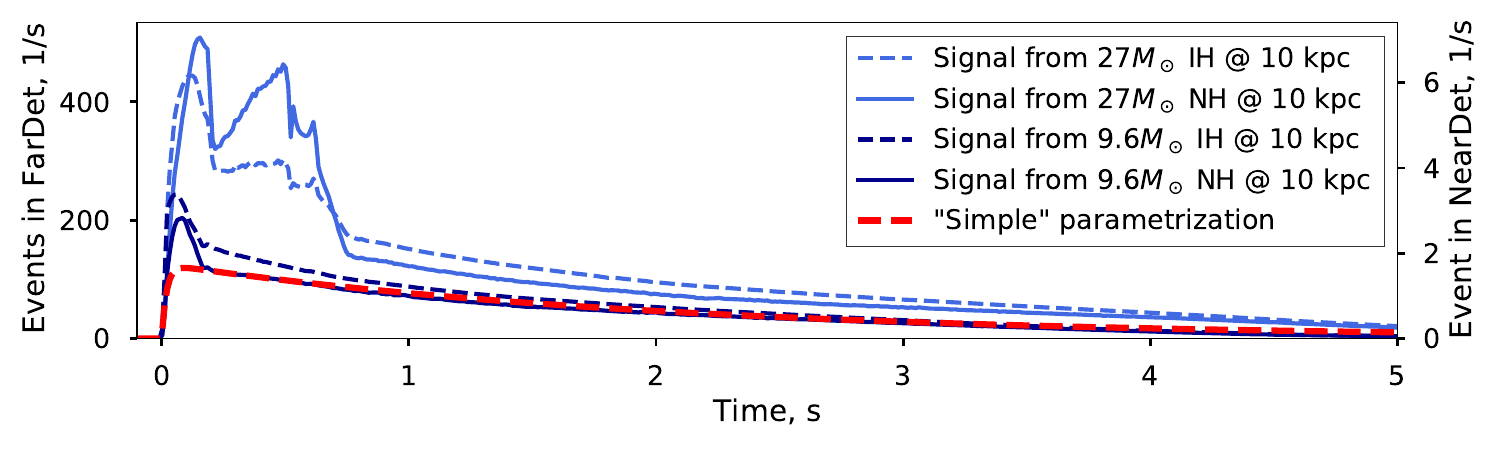}
    \caption{Expected signal shapes (number of neutrino interactions over the first five seconds of a supernova) in the NOvA detectors. The signal shapes are the same for FarDet and NearDet, and the rate is different by the ratio of detector masses (72.1). The signal rates are based on the Garching group simulations \cite{garching}  with two progenitor masses at \unit[10]{kpc} distance, assuming normal (NH) and inverted (IH) neutrino mass hierarchy. These simulations are compared to the analytical parametrization (see text for details).}
    \label{fig:nova_sg}
\end{figure}

NOvA is a long-baseline neutrino oscillation experiment which uses two liquid scintillator detectors: a small but low background near detector~(NearDet) underground, and a large far detector~(FarDet) on the surface with high background.
The selection procedure for the core-collapse supernova neutrino interaction candidates results in the background and signal rates shown in table~\ref{tab:nova_bg} and
figure~\ref{fig:nova_sg}. 
These expected signal event rates are obtained by multiplying the neutrino flux by the inverse beta-decay interaction cross-section \cite[Table~1]{StrumiaVissani}, the number of protons corresponding to the liquid scintillator amount in the NOvA detectors, and the detection efficiency from~\cite[figure~9, right]{nova_collaboration_supernova_2020}. 
Also we use a ''simple'' analytical parametrization of the signal by formula \eqref{eq:simple_param} with $\tau_0 = \unit[0.2]{s}, \tau_1=\unit[2]{s}$, which fits the exponential decay of the signal in time during the supernova cooling phase.
\begin{table}[htb]
    \centering
    \begin{tabular}{c|c|c}
         & Near detector &  Far detector \\
         \hline
        $N_{bg}$/s & 0.52 & 2483.21 \\
        $N_{sg}\,(\unit[9.6]{M_\odot})$ NH & 1.45 & 103.26  \\
        $N_{sg}\,(\unit[9.6]{M_\odot})$ IH & 1.81 & 130.34  \\
        $N_{sg}\,(\unit[27]{M_\odot})$  NH & 4.08 & 295.69 	\\
        $N_{sg}\,(\unit[27]{M_\odot})$  IH & 3.58 & 260.03 
    \end{tabular}
    \caption{Background and signal event numbers for the NOvA detectors during the first second of a simulated supernova. Signal rates are based on the Garching group simulations \cite{garching} with two progenitor masses at \unit[10]{kpc} distance, for both normal (NH) and inverted (IH) neutrino mass hierarchies.}
    \label{tab:nova_bg}
\end{table}

Using background values $B$ and selecting a signal model $S$ for equation~\eqref{eq:llr_calc} we can calculate the significance of observing the supernova neutrino burst, given the set of measured neutrino interactions with timestamps $\{t_i\}$. 
%Figure~\ref{fig:nova_timeseries} 
%\todo{which figure?} shows a simulated example of supernova neutrino burst (from $27 M_\odot$, NH model), observed in the NOvA detectors, and the significance calculated vs. assumed time of the supernova burst start (bottom plot). The significance, given by the counting analysis is given for comparison.

\subsection{Counting vs. shape analysis}

\begin{figure}[htb]
    \centering
    \includegraphics[scale=0.5]{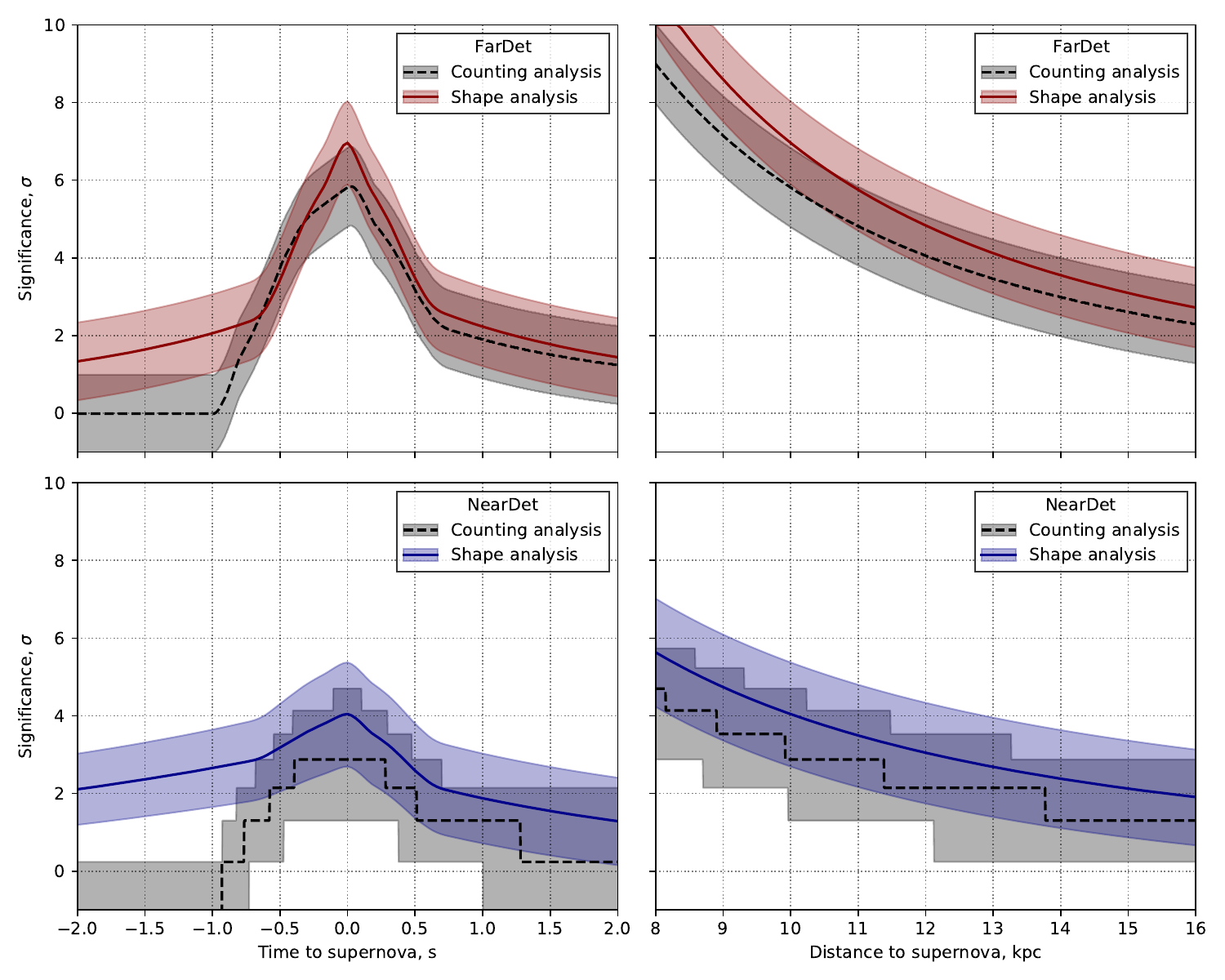}
    \caption{Comparison of significance obtained with CA and SA for a \unit[27]{$M_\odot$} supernova signal detection, assuming normal neutrino mass hierarchy, in the NOvA far (top panel) and near (bottom panel) detectors. Left: signal detection significance for a supernova at the \unit[10]{kpc} distance vs. the assumed time after the explosion start. Right: significance at the explosion starting time vs. distance to supernova. Lines show the mean value of the significance and the colored bands show the \unit[68]{\%} region.}
    \label{fig:nova_ca_sa}
\end{figure}

A comparison of the shape analysis performance with the standard counting analysis is made by calculating the significance for picking out the supernova signal above noise at various distances (figure~\ref{fig:nova_ca_sa}, right). The band shows the \unit[68]{\%} uncertainty assuming Poisson fluctuations of the background and signal event numbers.

The profile of the supernova significance vs. the assumed time of the SN start (figure~\ref{fig:nova_ca_sa}, left) shows that SA has maximum significance when the assumed time matches the actual start of the signal, while CA has a less pronounced peak, spread to as far as the counting analysis window (in case of low statistics, like in NearDet). This, and the symmetrical shape of the significance profile in the SA case allows a more precise determination of the SN burst start.

These advantages of SA over CA remain true (with a slight decrease in total significance) even in the case when expected signal model is different from the detected signal shape (as shown in~\cite{nova_collaboration_supernova_2020}). This is caused by the fact that different models still share the general features of the core-collapse signal.

\subsection{Expected signal model dependency}
The performance of the shape analysis method depends on the choice of the hypotheses $H_0$ and $H_1$, which are defined by the background $B(t)$ and signal $S(t)$ rates in \eqref{eq:llr_calc}. While background can be studied from the detector data (as in NOvA case), the signal shape is obtained from the supernova simulations, introducing a model uncertainty.

%To probe the dependency of the analysis performance on the expected signal we apply various analyses (CA and SA using various expected signal shapes shown on fig. \ref{fig:nova_sg}) to the list of received signals. Here we consider also the ``simple'' parametrization of the signal shape in order to demonstrate that even if an expected signal shape doesn't contain the detailed description of the real signal features, using an approximation in the SA provides much better results then using the CA.

Here we consider four signal models introduced in table~\ref{tab:nova_bg} as possible shapes of the actual supernova signal $S_{true}(t)$. 

We use \eqref{eq:pell_h_fourier} and (\ref{eq:pvalue} -- \ref{eq:trigger_eff}) to calculate the distribution of the significance value $z$ for the hypothesis, defined by an event rate $r(t) = B(t)+S_{true}(t)$, while using the same background $B(t)$ and selecting an expected signal model $S(t)$ for the definition of the test statistic function \eqref{eq:llr_calc}. This allows us to probe the analysis performance for various combinations of expected and input signal models $S(t)$ and $S_{true}(t)$.

Table \ref{tab:models_probe} summarizes the mean expected significance for the supernova signals at $d=\unit[10]{kpc}$ and the distance corresponding to $50\%$ efficiency, using the shape analysis with various signal hypotheses $S(t)$ (the four supernova models, ``simple'' parametrization), and counting analysis, vs. simulated input signal models $S_{true}(t)$. 

The table shows that the best results  in terms of the significance and detection distance are achieved when the received signal shape matches the expected one. However, even the ``simple'' parametrization reflecting only the most basic features of the core-collapse supernova signal shape provides a good improvement in comparison to the counting analysis.
\begin{table}[htb]
    \centering
     \def\arraystretch{1.2}
\begin{tabular}{lllrrrr}
\toprule
&\multicolumn{2}{c}{\multirow{2}{*}{Analysis}} & \multicolumn{4}{c}{Input model}\\
& &  &  $27M_\odot$ IH &  $27M_\odot$ NH &  $9.6M_\odot$ IH &  $9.6M_\odot$ NH \\
\midrule
\multirow{12}{*}{\rotatebox[origin=c]{90}{Distance ($\varepsilon=\unit[50]{\%}$), kpc}} &
\multirow{6}{*}{\rotatebox[origin=c]{90}{FarDet}} & CA &           10.13 &           10.80 &             7.17 &             6.38 \\
 &       & simple &  11.14 &          11.34 &           7.92 &             7.10 \\
 &       & $27M_\odot$ IH &   \textbf{11.27}&           11.62 &             7.96 &             7.14 \\
 &       & $27M_\odot$ NH &           11.13 &   \textbf{11.81}&             7.75 &             6.95 \\
 &       & $9.6M_\odot$ IH &           11.19 &          11.38 &     \textbf{8.02}&             7.16 \\
 &       & $9.6M_\odot$ NH &           11.21 &          11.40 &             8.01 &      \textbf{7.18}\\
\cline{2-7}
&\multirow{6}{*}{\rotatebox[origin=c]{90}{NearDet}} & CA &            7.08 &            7.56 &           5.03 &             4.50 \\
&        & simple &            8.56 &            8.62 &             6.10 &             5.49 \\
&        & $27M_\odot$ IH & \textbf{8.58} &             8.66 &             6.09 &             5.49 \\
&        & $27M_\odot$ NH &           8.50 &   \textbf{8.70} &             6.00 &             5.41 \\
&        & $9.6M_\odot$ IH &          8.50 &            8.57 &     \textbf{6.10}&             5.48 \\
&        & $9.6M_\odot$ NH &          8.49 &            8.58 &             6.08 &     \textbf{5.47}\\
\bottomrule
\toprule

\multirow{12}{*}{\rotatebox[origin=c]{90}{$z_{mean}$ at \unit[10]{kpc}, $\sigma$}} & 
\multirow{6}{*}{\rotatebox[origin=c]{90}{FarDet}} & CA &            5.12 &            5.82 &             2.58 &             2.04 \\
&        & simple &            6.01 &            6.23 &             2.87 &             2.23 \\
&        & $27M_\odot$ IH &     \textbf{6.24}&            6.64 &             3.04 &             2.41 \\
&        & $27M_\odot$ NH &            6.09 &     \textbf{6.87}&             2.88 &             2.28 \\
&        & $9.6M_\odot$ IH &            6.08 &            6.30 &     \textbf{3.00}&             2.33 \\
&        & $9.6M_\odot$ NH &            6.08 &            6.30 &             2.95 &     \textbf{2.30}\\
\cline{2-7}
&\multirow{6}{*}{\rotatebox[origin=c]{90}{NearDet}} & CA &            2.88 &            2.88 &             1.30 &             1.30 \\
&        & simple &            3.93 &            3.98 &             2.27 &             1.90 \\
&        & $27M_\odot$ IH &     \textbf{3.95}&            4.01 &             2.26 &             1.90 \\
&        & $27M_\odot$ NH &            3.91 &     \textbf{4.05}&             2.24 &             1.88 \\
&        & $9.6M_\odot$ IH &            3.90 &            3.96 &     \textbf{2.29}&             1.92 \\
&        & $9.6M_\odot$ NH &            3.90 &            3.96 &             2.29 &     \textbf{1.92}\\
\bottomrule
\end{tabular}

    \caption{Performance metrics of the analyses for various expected signal shapes (rows) vs. input signal models $S_{true}(t)$ (columns): supernova distance, where the efficiency is \unit[50]{\%} (top) and average expected significance for the supernova at \unit[10]{kpc} (bottom). 
    The counting analysis is performed in the \unit[1]{second} time window, and the shape analysis uses expected signals for the supernovae at \unit[10]{kpc} distance. The bold numbers mark the case when the expected) signal matches the received one.}
    \label{tab:models_probe}
\end{table}

\subsection{Estimating supernova signal start time}

One of the tasks of the supernova detection system is the reconstruction of the supernova signal start time $t^*_{rec}$ for the given set of data $\{t_i\}$.

As the significance $z(t^*,\{t_i\})$ measures a goodness of fitting the data with the supernova signal starting at the time $t^*$, we use the time of maximal significance as an estimator for $t^*$:
\begin{equation}
    t_{rec}^* = \operatornamewithlimits{argmax}_{t^*} z(t^*, \{t_i\}).
    \label{eq:trec}
\end{equation}

\begin{figure}[!htb]
    \centering
    \includegraphics[scale=0.5]{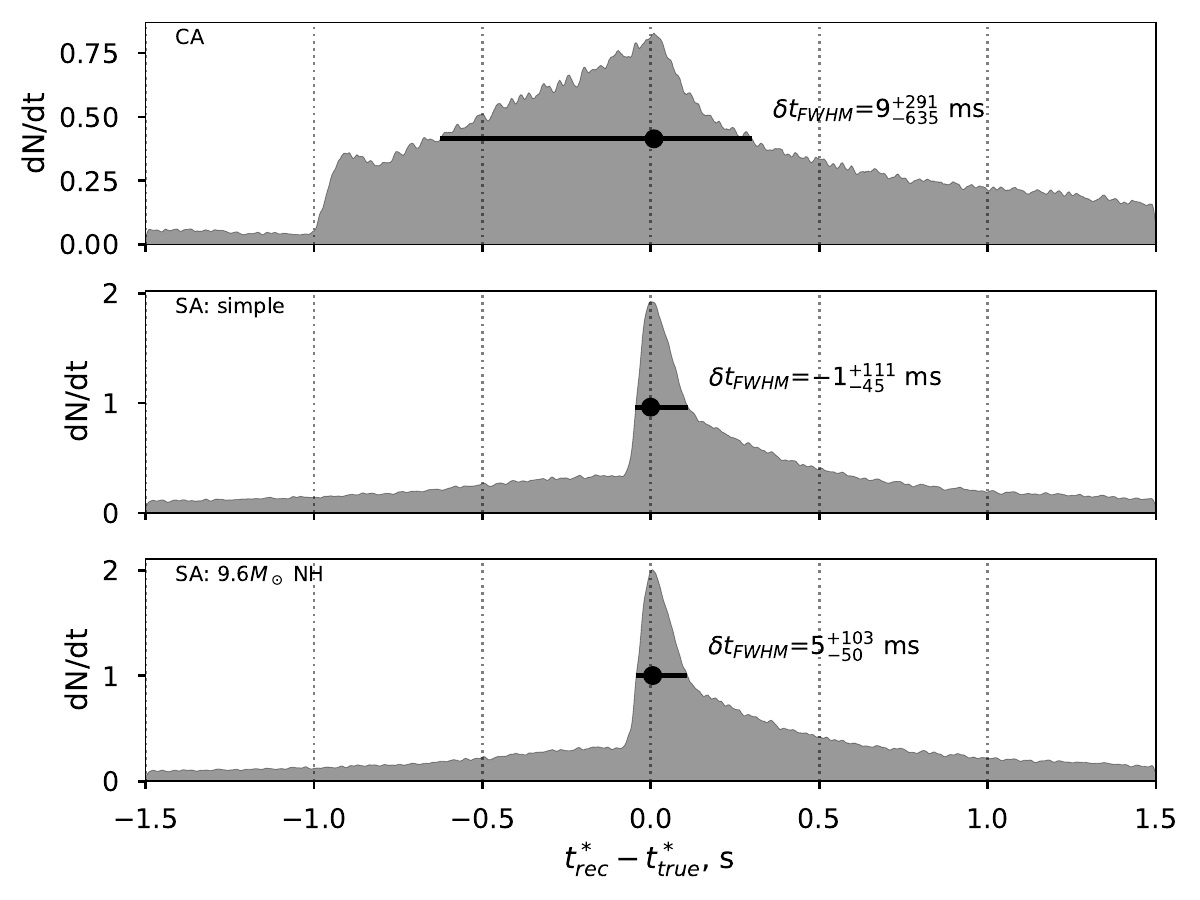}
    \caption{Distribution of the supernova start time estimation error $t^*_{rec}-t^*_{true}$ for the simulated samples of neutrino interactions from $\unit[9.6]{M_\odot}$~NH at $d=\unit[10]{kpc}$ distance in the NOvA far detector. The panels show results of CA (top), SA with ``simple'' parametrization (middle) and SA with the signal shape corresponding to the injected signal (bottom). The numbers show the distribution spread, estimated using the full width at half maximum (FWHM) method. }
    \label{fig:nova_timing}
\end{figure}

In order to study the precision of such estimator, we perform a large number of toy Monte-Carlo simulations of the neutrino event timestamps $\{t_i\}$, 
with each neutrino time drawn from a distribution following the event rate $R(t) = B + S_{true}(t)$, where $S_{true}(t)$ is one of the input signals from $d=\unit[10]{kpc}$ distant supernova starting at a given time $t^*_{true}$. 
Then we analyze these simulated datasets assuming various signal hypotheses in the same way as in previous section, calculating the $t^*_{rec} - t^*_{true}$. 
Figure \ref{fig:nova_timing} shows such distributions for the simulated signal from $\unit[9.6]{M_\odot}$~NH signal in the NOvA far detector, processed by the three different analyses. 
Since the tails of the distribution of the $t_{rec}$ defined in \eqref{eq:trec} depend on the considered time window, one cannot use the standard deviation or the distribution quantiles in order to evaluate the distribution width (and thus $t_{rec}^*$ estimator precision).
Instead we use the time of the distribution maximum and the full width on the half maximum (FWHM) of the distribution peak as a metric of the method precision. 

The time precision estimation procedure has been applied to all combinations of expected and received signals as in the layout of table~\ref{tab:models_probe}. Typical FWHM values of the obtained distributions accord with the example depicted in figure~\ref{fig:nova_timing} ($\sim\unit[1000]{ms}$ for CA and $\sim\unit[200]{ms}$ for SA) showing a robust $\sim5$~times improvement in timing precision when accounting for the signal shape.

%Table \ref{tab:nova_timing} show the same metrics for the method precision for all considered analyses and simulated signal shapes.

%\begin{table}[!htb]
%    \centering
%     \include{fig/nova/table_timing}
%    \caption{Error of the supernova start time estimation $t^*_{rec}-t^*_{true}$ in milliseconds, for various analyses and received signals from $d=\unit[10]{kpc}$ supernovae. Large number show the most probable error value (maximum of the distribution), and plus and minus values show the limits of the FWHM of the distribution. The bold numbers are for the case when expected signal matches to the received one.}
%    \label{tab:nova_timing}
%\end{table}

\subsection{Results}

\begin{figure}[!htb]
    \centering
    \includegraphics[scale=0.5]{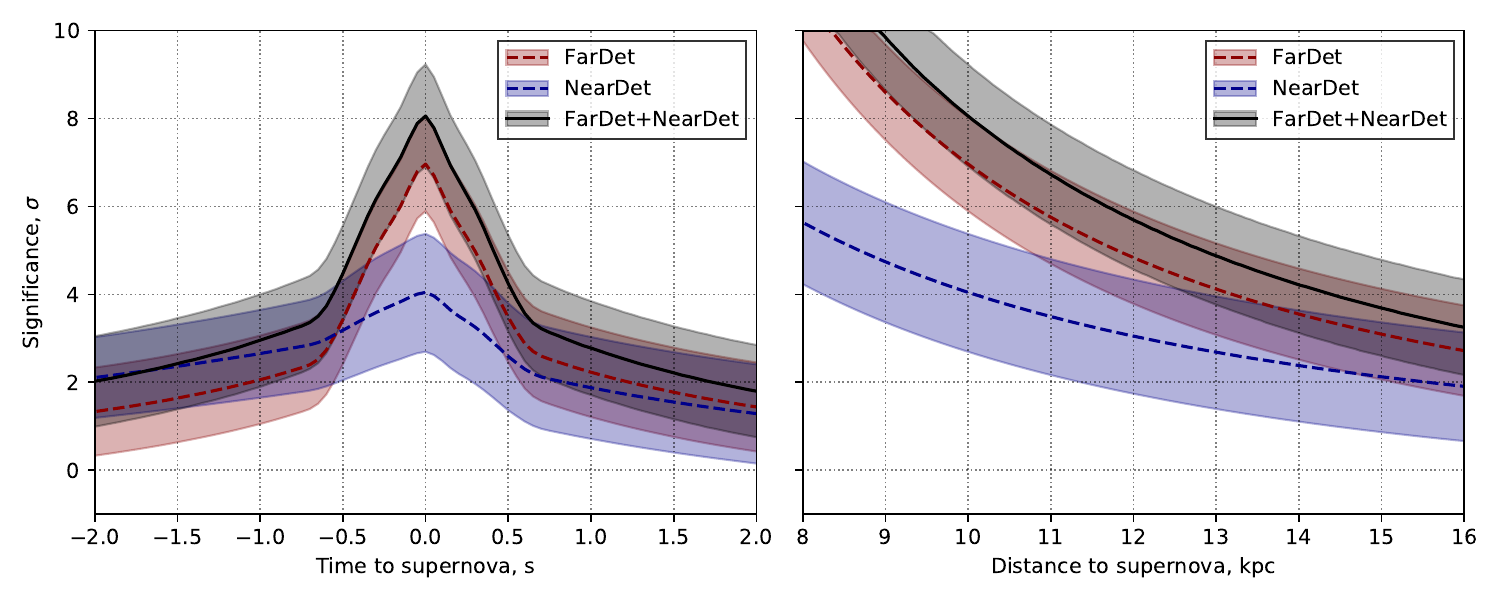}
    \caption{Significance of the supernova signal detection for the NOvA near and far detectors and their combination. Layout is the same as in figure~\ref{fig:nova_ca_sa}.}
    \label{fig:nova_combined}
\end{figure}

Currently the triggering systems on NOvA detectors work independently of each other.
However combining signals from NearDet and FarDet can improve the total sensitivity of the system, as shown in figure~\ref{fig:nova_combined}.

Possible time shifts between the detectors might affect the combined significance, however most experiments use GPS time synchronization with the uncertainty below \unit[100]{ns}\cite{Minos_timing}, which is much lower than the scale of the significance peaks on figure \ref{fig:nova_combined}. Moreover, since the detection is based on pattern matching rather than determination of the sharp rise of the signal front, these shifts also have negligible impact to the precision of the determination of $t^*_{rec}$.

Table~\ref{tab:nova_results} summarizes the results of the calculation of the expected NOvA reach, assuming detection threshold $z_{thr}=\unit[5]{\sigma}$ and detection efficiency $\varepsilon=\unit[50]{\%}$, for various signal models and depending on the applied analysis. This calculation shows that using the shape analysis for individual detectors allows detection about \unit[1]{kpc} further than counting analysis. An additional increase of about \unit[1]{kpc} is achieved by applying the combined shape analysis.
\begin{table}[!htb]
    \centering
    \begin{tabular}{llrrrrr}
\toprule
Metric & Model& \multicolumn{2}{l}{Near detector} & \multicolumn{2}{l}{Far detector} & Far+Near \\
     &                 & CA & SA & CA & SA & Joint SA \\
\midrule
Distance ($\varepsilon=\unit[50]{\%}$), kpc & 27$M_\odot$ IH &              7.08 &           8.58 &             10.13 &          11.27 &          12.36 \\
     & 27$M_\odot$ NH &              7.56 &           8.70 &             10.80  &          11.81 &          12.85 \\
     & 9.6$M_\odot$ IH &              5.03 &           6.10 &              7.17 &           8.02 &           8.80 \\
     & 9.6$M_\odot$ NH &              4.50 &           5.47 &              6.38 &           7.18 &           7.89 \\
\midrule
$z_{mean}$ at \unit[10]{kpc}, \unit{$\sigma$} & 27$M_\odot$ IH &              2.88 &           3.95 &              5.12 &           6.24 &           7.47 \\
     & 27$M_\odot$ NH &              2.88 &           4.05 &              5.82 &           6.87 &           8.06 \\
     & 9.6$M_\odot$ IH &              1.30 &           2.29 &              2.58 &           3.00 &           3.95 \\
     & 9.6$M_\odot$ NH &              1.30 &           1.92 &              2.04 &           2.30 &           3.20 \\
\bottomrule
\end{tabular}

    \caption{Comparison of the distance and significance metrics using CA and SA for individual detectors, and then using a combined shape analysis (Joint SA). Metrics are: maximal supernova distance, where the detection efficiency exceeds \unit[50]{\%} (top part), and value for $z_{mean}$ expected for supernova distance \unit[10]{kpc} (bottom part). The SA results are shown for the case when expected and input signal models match. Models correspond to various progenitor star masses and neutrino mass hierarchies, as listed in table~\ref{tab:nova_bg}.}
    \label{tab:nova_results}
\end{table}

\newpage
\section{Pre-supernova neutrino detection}
\label{seq:presn}

Presupernova neutrinos are emitted in the late stages of stellar evolution as the degenerate core of the star becomes hot enough that the dominant cooling mechanism is neutrino emission, producing a flux which gradually increases with time for several days up to the moment of the core collapse.

To probe how the choice of models might effect the calculations, we consider several predictions of the presupernova neutrino flux: the calculations by Odrzywolek~\cite{Odrzywolek:2010zz}, Patton et al.~\cite{patton_neutrinos_2017} and Kato et al.~\cite{kato_neutrino_2017} for a $\unit[15]{M_{\odot}}$ progenitor star.

\begin{figure}[!htb]
    \centering
    \includegraphics[scale=0.5]{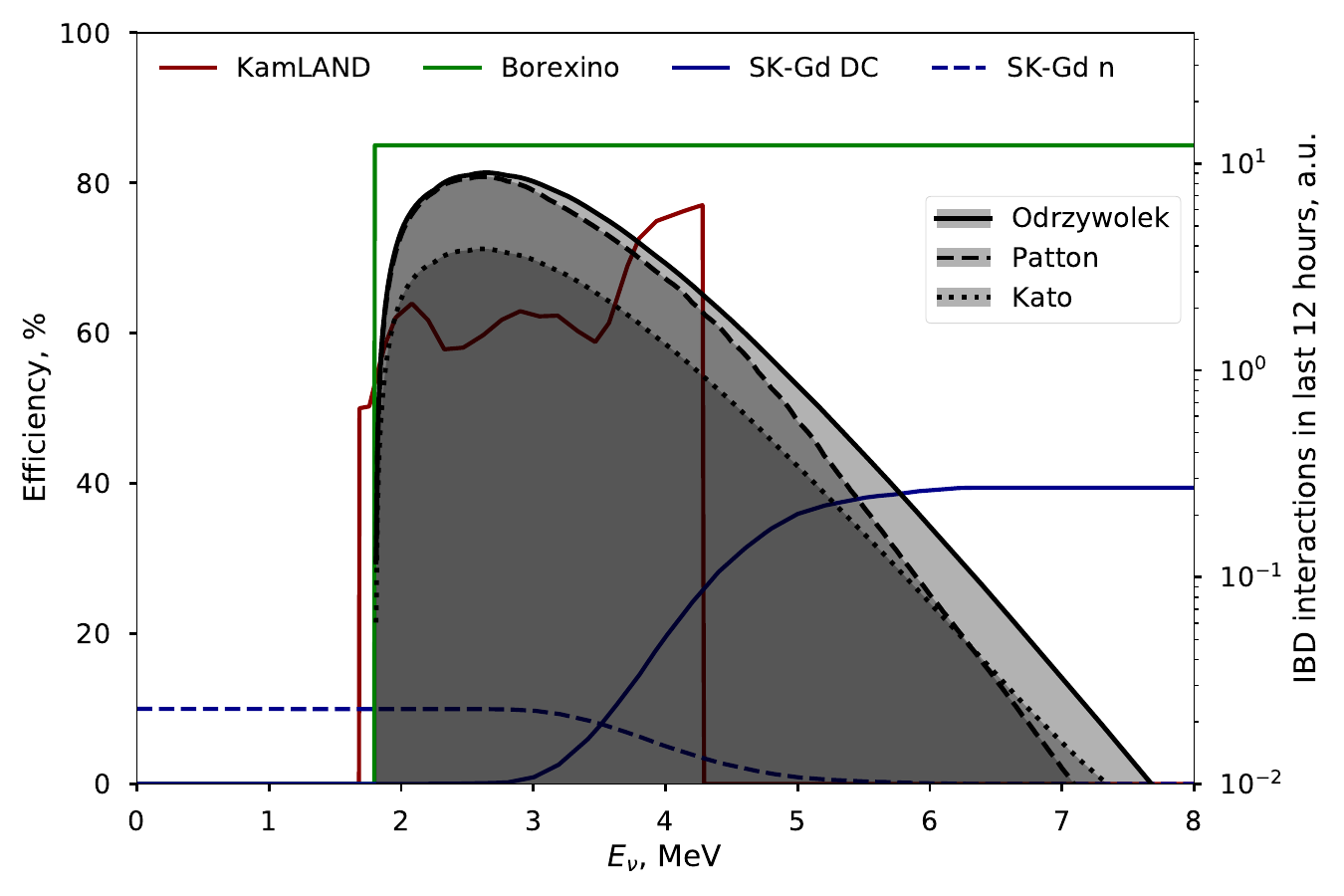}
    \caption{IBD detection efficiencies vs. $\bar{\nu}_e$ neutrino energy for various experiments and detection modes (solid and dashed lines), compared to the expected $\bar{\nu}_e$ spectra in last 12 hours before the supernova for the three models considered (shaded regions).}
    \label{fig:presn_eff}
\end{figure}

\subsection{Detectors}
As in the case of the core-collapse supernova neutrino burst, the presupernova neutrinos can be detected by various neutrino and dark matter experiments in several interaction channels~\cite{kato_theoretical_2020}. Here we will limit our consideration to inverse beta-decay (IBD) channel in three experiments: KamLAND, Borexino and SK-Gd.

\subsubsection*{KamLAND}
KamLAND~\cite{KamLAND:2002uet} is a \unit[1]{kiloton} liquid scintillator detector, located in the Kamioka mine. KamLAND detects the low-energy IBD interactions as a delayed coincidence of positron annihilation and neutron capture signals, and applies series of cuts and a likelihood variable based on energy and position to select the signal interactions and suppress background. For this work we use the efficiency from the KamLAND analysis from \cite[figure~4 (top)]{asakura_kamland_2016}, and apply the final selection on the positron energy $\unit[0.9]{MeV}\leq E_p\leq\unit[3.5]{MeV}$. The energy resolution of $\unit[6.4]{\%}/\sqrt{E(\unit{MeV})}$ would have little effect on the resulting efficiency, so in this work we assume $E_p = E_{\nu} - \unit[0.87]{MeV}$. KamLAND searches for the presupernova signal using a counting analysis with a \unit[48]{hour} time window.

\subsubsection*{Borexino}
Borexino~\cite{Borexino:2008gab} uses a high radio-purity liquid scintillator detector, located in the Gran Sasso underground laboratory, which provides a very low radioactive and cosmic background for the anti-neutrinos in the considered energy range. The Borexino collaboration performed a search for anti-neutrino signal from the diffuse supernova background~\cite{agostini_search_2019}, and the same analysis can be applied to the detection of the pre-supernova IBD signals. The IBD detection efficiency in Borexino is dominated by the neutron capture detection and is estimated to be \unit[85]{\%}. However, a relatively small fiducial mass of \unit[220]{t} limits the Borexino sensitivity to weak signals such as presupernova neutrinos.
Nevertheless, including Borexino in combination with other detectors can improve the overall presupernova signal sensitivity and supernova prediction time. For a Borexino presupernova counting analysis we use a time window of \unit[48]{hours}, the same as in KamLAND.

\subsubsection*{SK-Gd}
SK-Gd~\cite{simpson_sensitivity_2019} is the next phase of the \unit[50]{kt} water Cherenkov detector Super-Kamiokande. The addition of Gadolinium enables the detection of the neutron capture, thus the IBD interactions can be seen as a delayed coincidence (DC) of the positron and neutron signals, or just the neutron capture (n) if the positron signal was not reconstructed.  
These two modes can be considered as two independent detection channels, and their combination illustrates the use of a joint analysis within the same experiment.

For this paper the {DC}-mode efficiency is taken from~\cite[figure~8, black curve]{simpson_sensitivity_2019}, and the neutron capture efficiency is taken as \unit[10]{\%}, considering the optimistic Gd capture model. Additionally, in order to make {SK-Gd DC} and {neutron} modes independent, we multiply the neutron capture efficiency by the probability of not detecting the positron signal $(1-\varepsilon_{pos}(E_\nu))$. SK-Gd has a higher background rate than KamLAND and Borexino, therefore the optimal time window for counting analysis is smaller: \unit[12]{hours} for both detection modes.

\subsection*{Expected signals and backgrounds}

\begin{figure}[htb]
    \centering
    \includegraphics[scale=0.5]{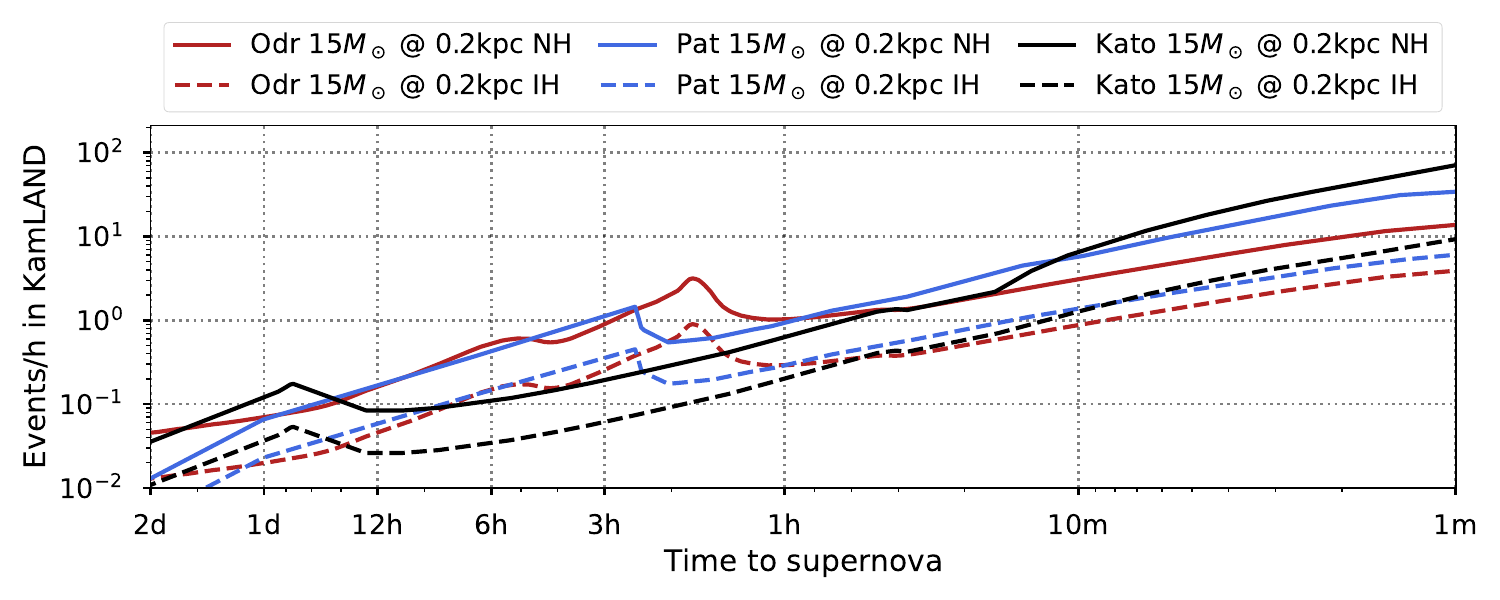}
    \caption{Expected event rates from a presupernova signal in the KamLAND detector. Three flux models and two neutrino mass hierarchies are considered for a progenitor mass $M_\odot=\unit[15]{MeV}$ at a \unit[200]{pc} distance.}
    \label{fig:presn_sigshape}
\end{figure}

The detection efficiencies of the considered detectors vs. anti-neutrino energy $E_\nu$, as well as presupernova anti-neutrino energy spectra in the considered models, are depicted in figure~\ref{fig:presn_eff}. 
Figure~\ref{fig:presn_sigshape} shows the resulting presupernova event rates versus time expected in the KamLAND experiment for the models considered in this work.

Table~\ref{tab:presn_detector_rates} summarizes the signal event rates expected for different presupernova models, background levels, and the time windows chosen for the counting analysis, for each experiment considered. 

\begin{table}[htb]
\centering
\begin{tabular}{llllll}
\toprule
\multirow{2}{*}{Detector} & \multicolumn{3}{c}{$N_{sg}$ in the last hour before SN}  & \multirow{2}{*}{$N_{bg}$/hour} & \multirow{2}{*}{Counting window}\\
        &               Kato15 &                Odr15 &                Pat15 &   & \\
\midrule
%JUNO       &       240 (    39.4) &      84.5 (      24) &       153 (    34.6) &         0.7 & \unit[24]{h} \\
Borexino   &      3.95 (0.705) &      1.15 (0.327) &      2.11 (0.5) &      0.0014 & \unit[48]{hours} \\
KamLAND    &      7.13 (1.19) &       2.4 (0.681) &      4.39 (1.0) &      0.0029 & \unit[48]{hours}\\
%SK-Gd\_DC\_0 &      80.4 (    16.2) &      14.6 (    4.14) &      27.7 (    7.57) &           1 \\
%SK-Gd\_n\_0  &      30.8 (    5.11) &      10.8 (    3.06) &      19.5 (    4.42) &         5.5 \\
SK-Gd DC &      86.6 (17.5) &      15.7 (4.45) &      29.7 (8.13) &           1 &\unit[12]{hours} \\
SK-Gd neutron  &      42.5 (7.05) &      14.8 (4.21) &      26.8 (6.09) &         5.5 & \unit[12]{hours} \\
\bottomrule
\end{tabular}
\caption{Presupernova neutrino event rates during the last hour before the supernova for the three neutrino flux models at a \unit[200]{pc} distance  and normal (inverted) neutrino mass hierarchy; the background conditions; the counting analysis time window.}
\label{tab:presn_detector_rates}
\end{table}

\subsection{Counting vs. shape analysis}

To demonstrate features of the shape analysis applied to the presupernova neutrino search, we hereafter consider the Odrzywolek model and normal neutrino hierarchy, unless noted otherwise.

While the detection of the core collapse supernova neutrino signal would provide the precise timing and other physical information of the supernova explosion process, the presupernova signal has an additional goal of predicting the supernova event beforehand. Thus the metric of how far in advance we can detect the supernova is as important as the maximal detection distance.

The shape analysis probes how well the signal+background shape fits the acquired data, so the optimal discrimination (maximal significance) is reached when the full-length signal is present in the data. However, since the signal shapes in figure~\ref{fig:presn_sigshape} show almost monotonic behaviour in time, even an incomplete early part of the signal in the data can provide a considerable significance elevation when fitted with the full signal shape. % resulting in a more efficient prediction of a supernova explosion.

\begin{figure}[htb]
    \centering
    \includegraphics[scale=0.45]{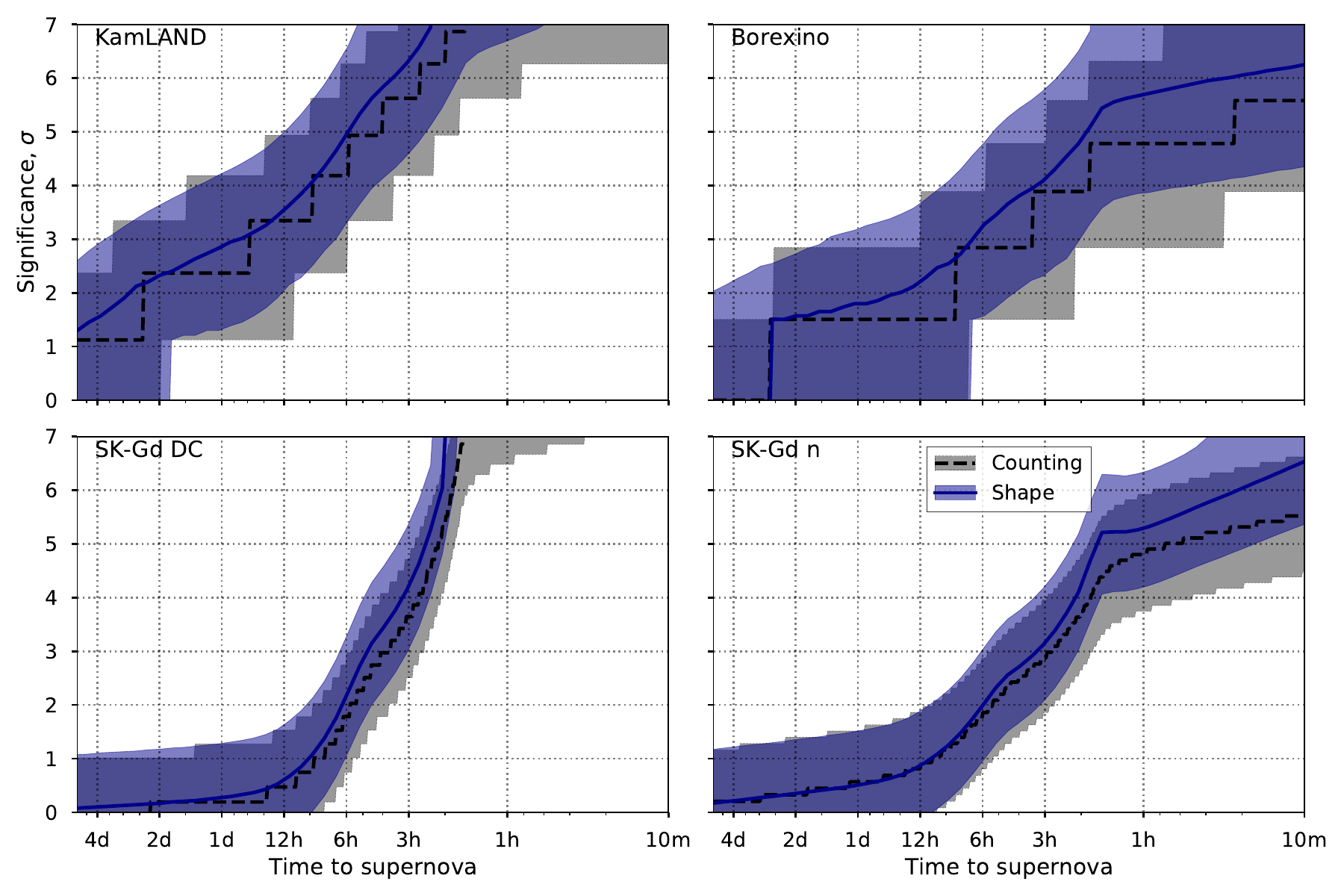}
    \caption{Expected significance for the various detectors using counting (dashed line) and shape (solid line) analyses for the Odrzywolek NH model at \unit[200]{pc} distance vs. the time to supernova. The filled regions show the \unit[68]{\%} band of significance value.}
    \label{fig:presn_zs_time}
\end{figure}

\begin{figure}[htb]
    \centering
    \includegraphics[scale=0.45]{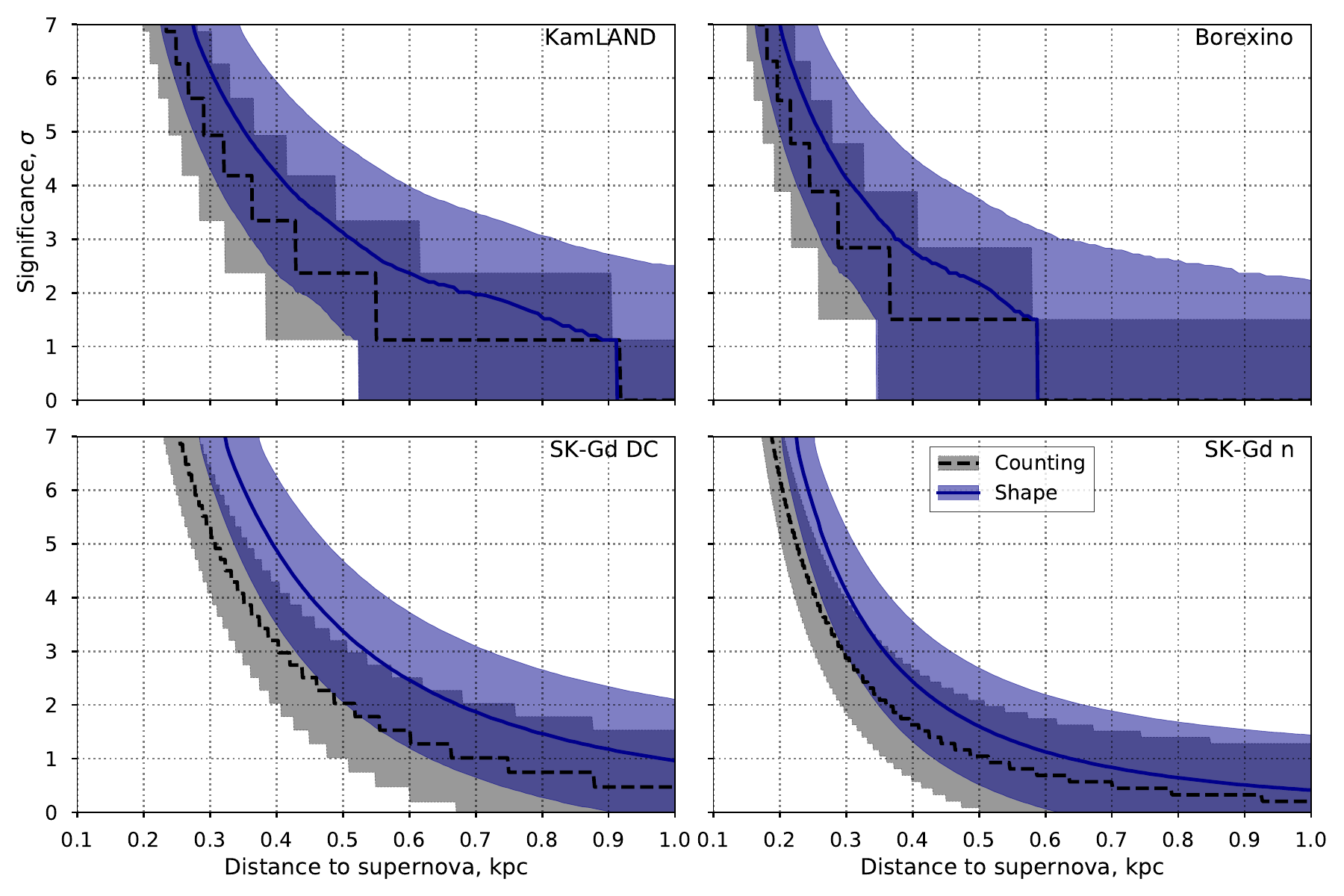}
    \caption{Expected significance for the various detectors using counting (dashed line) and shape (solid line) analyses for the Odrzywolek NH model 1~minute before the supernova explosion vs. the distance to supernova. The filled regions show the \unit[68]{\%} band of significance value.}
    \label{fig:presn_zs_dist}
\end{figure}

In other words, applying the shape analysis improves significance even at the early stages of the presupernova signal. Figure~\ref{fig:presn_zs_time} demonstrates that the shape analysis allows an earlier detection of the presupernova signal from \unit[200]{pc} distance 
%(Odrzywolek model, NH) 
than the counting analysis. The panels in figure~\ref{fig:presn_zs_dist} show the dependency of the maximal presupernova significance (\unit[1]{minute} prior to the collapse) on the supernova distance, for the same model.

\begin{figure}[htb]
    \centering
    \includegraphics[scale=0.45]{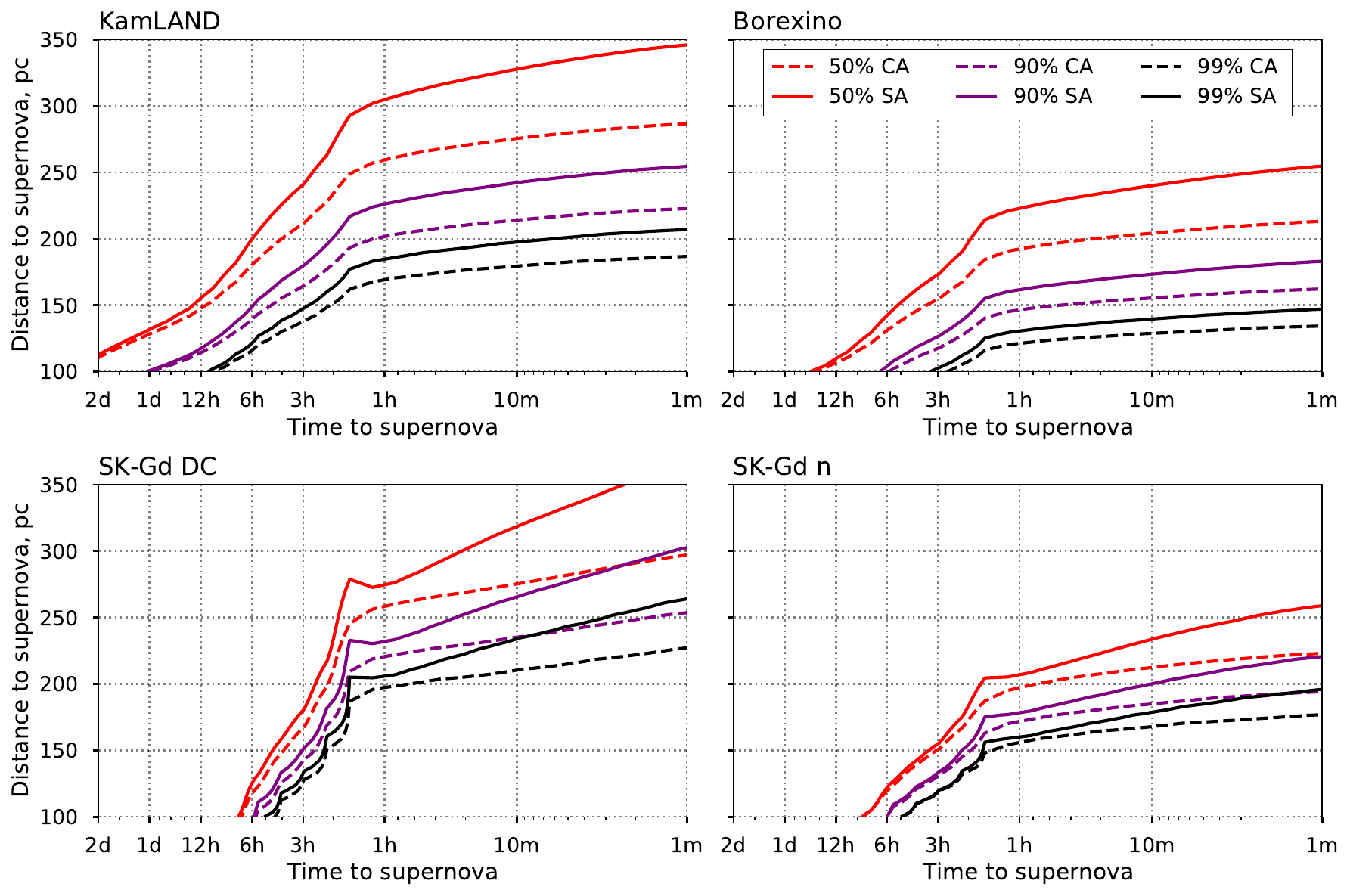}
    \caption{Presupernova detection reach vs. time to supernova and supernova distance with \unit[50]{\%}, \unit[90]{\%} and \unit[99]{\%} efficiency. The solid and dashed lines show the results of shape and counting analyses, respectively. Odrzywolek NH model is assumed for both expected and received signals.}
    \label{fig:2d_casa_reaches}
\end{figure}

Figure~\ref{fig:2d_casa_reaches} shows the regions where the detection efficiency is above \unit[50]{\%}, \unit[90]{\%} and \unit[99]{\%} in the counting and shape analyses for the Odrzywolek model with NH, assuming the detection threshold $z_{thr}=\unit[5]{\sigma}$. The exact value of this threshold in real detection system should be defined by the tolerable false alarm probability. 

\subsection{Results}
\begin{figure}[htb]
    \centering
    \includegraphics[scale=0.5]{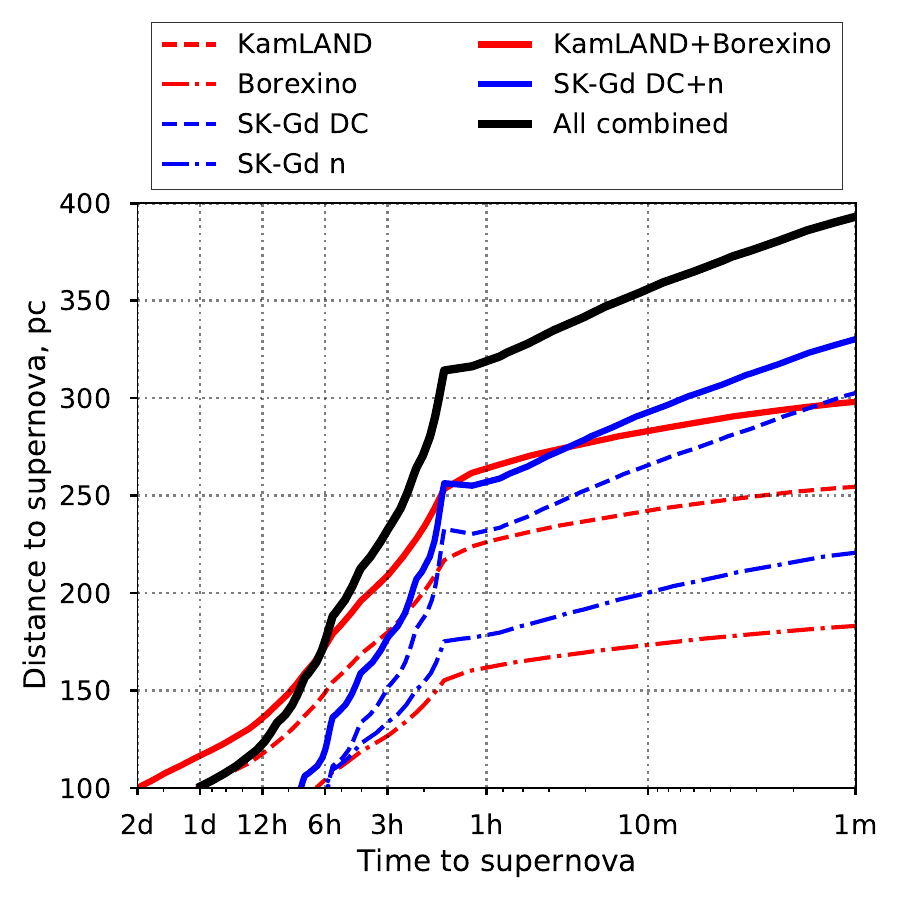}
    \caption{Presupernova detection reach with \unit[90]{\%} efficiency for the shape analyses for the individual detectors (dashed, dashdot lines) and their combinations (solid lines).  Odrzywolek NH model is assumed for both expected and received signals.}
    \label{fig:2d_combine_reaches}
\end{figure}

%\subsection{Dependency on the expected signal model}
Joint analysis of the detectors using ~\eqref{eq:pell_comb} provides an additional improvement of the presupernova signal detection efficiency. Figure~\ref{fig:2d_combine_reaches} shows the regions with \unit[90]{\%} efficiency for separate shape analyses and their combinations. Including of the analyses with low sensitivity to the presupernova signal, such as Borexino or SK-Gd neutron mode, still increases the combined reach. For example, the \unit[90]{\%} efficiency of detection of a presupernova signal at \unit[200]{pc} distance occurs about two hours earlier for Borexino+KamLAND, than only for KamLAND.
However the combination doesn't provide large benefits at the very early stages ($>\unit[6]{hours}$ before a supernova), because the shape analysis is optimized on fitting the full signal shape.

\begin{figure}[htb]
    \centering
    \includegraphics[scale=0.5]{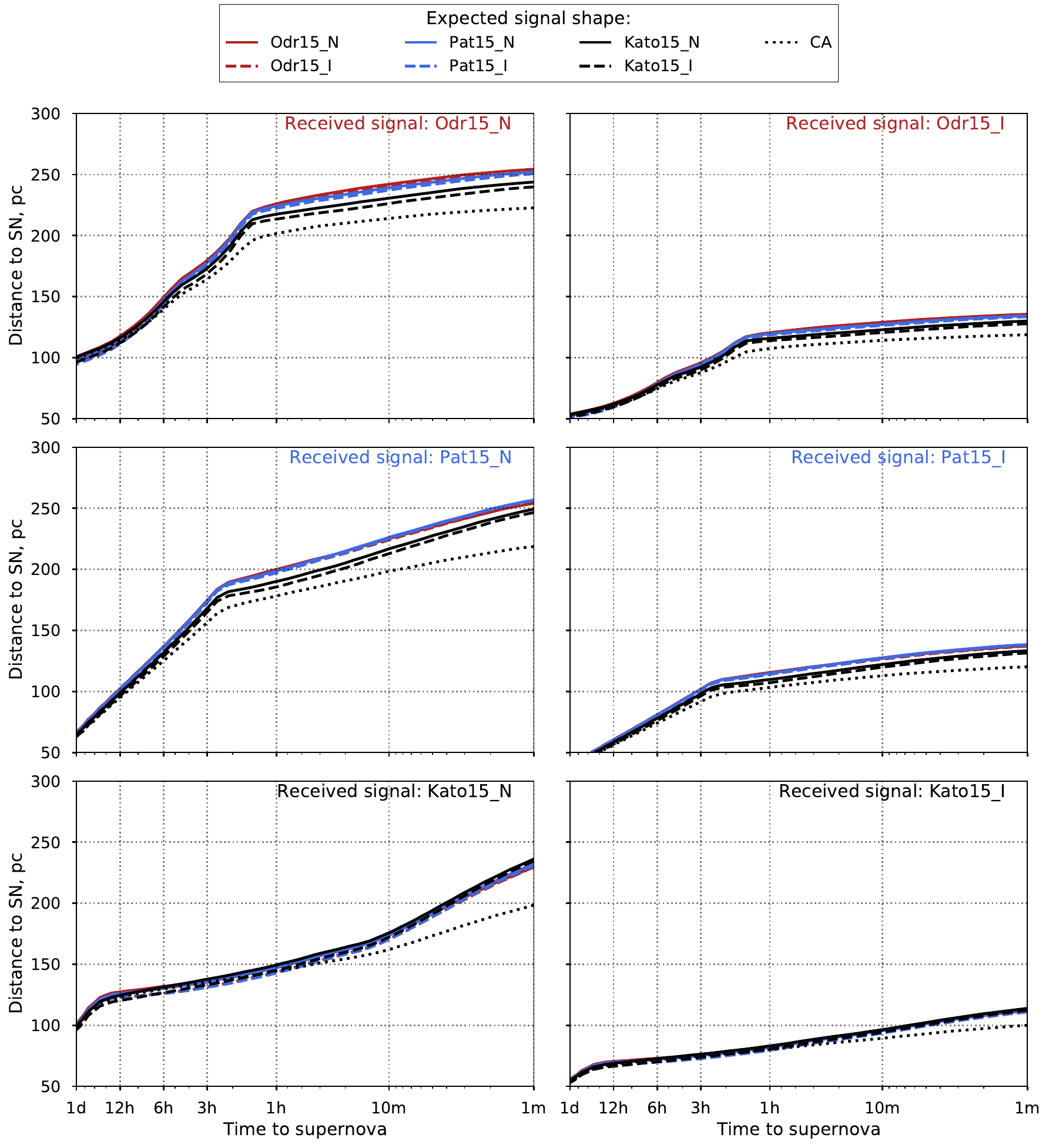}
    \caption{Presupernova detection reach with \unit[90]{\%} efficiency for the various received signals, using the shape analyses with various expected signals (solid and dashed lines) and the counting analysis (dotted lines)}
    \label{fig:reaches_for_various_models}
\end{figure}

The accuracy and performance of the shape analysis method depends on the knowledge of the background and signal rates. Since the actual signal is not known in advance, we study the uncertainty of the method by calculating the sensitivity reaches for various signal models and compare them to the counting analysis. Figure~\ref{fig:reaches_for_various_models} shows that while the maximal reach is always obtained when the expected signal and the received signal coincide, other models still show better reach than the counting analysis. This reflects the fact that even a wrong signal model describes the presupernova signal better than a constant rate in a given time window (as in the counting analysis).

Table \ref{tab:presn_results} presents quantitative metrics of the expected reach, prediction time and significance using various analysis methods for the detection modes in considered detectors and their combinations. The performance was calculated for all above-mentioned presupernova models and neutrino mass hierarchies using the Odrzywolek NH model as the expected signal in SA.
Distance and time in this table correspond to the \unit[50]{\%} efficiency detection reach (in contrast to \unit[90]{\%} in figures~\ref{fig:2d_combine_reaches},~\ref{fig:reaches_for_various_models}). 

\begin{table}[htb]
    \centering
    \begin{tabular}{lllrrrrrr}
\toprule
     &                                       &       & \multicolumn{2}{c}{Kato15} & \multicolumn{2}{c}{Odr15} & \multicolumn{2}{c}{Pat15} \\
& Experiment & Analysis&      IH &      NH &      IH &      NH &      IH &      NH \\

\multirow{11}{*}{\rotatebox[origin=c]{90}{$d(50\%)$, pc}} & Borexino & CA &  93.8 & 180.9 & 112.5 & 211.1 & 112.9 & 203.4 \\
     &                                       & SA & 108.5 & 216.2 & 134.0 & 251.4 & 134.1 & 245.9 \\
     & KamLAND & CA & 125.4 & 242.7 & 151.5 & 284.1 & 152.5 & 274.9 \\
     &                                       & SA & 146.2 & 293.1 & 182.1 & 341.6 & 183.3 & 336.2 \\
     & SK-Gd DC & CA & 122.3 & 236.2 & 155.2 & 291.1 & 150.5 & 272.1 \\
     &                                       & SA & 170.9 & 345.0 & 188.8 & 354.3 & 180.5 & 336.8 \\
     & SK-Gd n & CA &  82.4 & 167.4 & 117.7 & 220.8 & 119.4 & 218.8 \\
     &                                       & SA & 120.4 & 265.0 & 146.0 & 273.9 & 145.9 & 282.6 \\
     & Borexino+KamLAND & Joint & 169.9 & 339.9 & 211.0 & 395.9 & 212.0 & 388.9 \\
     & SK-Gd DC+n & Joint & 181.3 & 375.5 & 206.0 & 386.5 & 199.8 & 377.9 \\
     & All combined & Joint & 214.5 & 439.8 & 250.9 & 471.0 & 246.6 & 461.0 \\
\midrule
\multirow{11}{*}{\rotatebox[origin=c]{90}{$t(50\%)$, minutes}} & Borexino & CA &      - &      - &      - &  20.67 &      - &   1.92 \\
     &                                       & SA &      - &   2.48 &      - & 112.80 &      - &  39.83 \\
     & KamLAND & CA &      - &  16.89 &      - & 239.25 &      - & 183.18 \\
     &                                       & SA &      - &  47.75 &      - & 355.11 &      - & 285.07 \\
     & SK-Gd DC & CA &      - &   3.04 &      - & 127.69 &      - &  51.05 \\
     &                                       & SA &      - &   8.57 &      - & 149.22 &      - & 130.97 \\
     & SK-Gd n & CA &      - &      - &      - &  46.73 &      - &   9.25 \\
     &                                       & SA &      - &   5.39 &      - & 118.03 &      - &  61.46 \\
     & Borexino+KamLAND & Joint &      - & 280.89 &  12.82 & 500.34 &   7.39 & 408.41 \\
     & SK-Gd DC+n & Joint &      - &  13.65 &   2.40 & 212.30 &      - & 199.52 \\
     & All combined & Joint &   2.08 &  84.13 & 100.66 & 431.33 &  26.53 & 379.01 \\
     \midrule
\multirow{11}{*}{\rotatebox[origin=c]{90}{$z_{mean}(\unit[200]{pc})$, \unit{$\sigma$}}} & Borexino & CA &   1.51 &   4.80 &   1.51 &   5.60 &   1.51 &   5.60 \\
     &                                       & SA &   2.39 &   6.12 &   2.95 &   6.76 &   3.01 &   6.74 \\
     & KamLAND & CA &   2.37 &   7.45 &   3.35 &   8.51 &   3.35 &   8.51 \\
     &                                       & SA &   3.45 &    $>10$ &   4.50 &    $>10$ &   4.56 &    $>10$ \\
     & SK-Gd DC & CA &   2.51 &   8.29 &   3.20 &   9.81 &   3.20 &   9.15 \\
     &                                       & SA &   4.85 &    $>10$ &   4.84 &    $>10$ &   4.72 &    $>10$ \\
     & SK-Gd n & CA &   0.93 &   4.07 &   1.86 &   6.03 &   1.86 &   6.13 \\
     &                                       & SA &   2.10 &  $>10$ &   2.79 &   9.82 &   2.85 &  $>10$ \\
     & Borexino+KamLAND & Joint &   4.25 &    $>10$ &   5.45 &    $>10$ &   5.53 &    $>10$ \\
     & SK-Gd DC+n & Joint &   5.21 &    $>10$ &   5.57 &    $>10$ &   5.50 &    $>10$ \\
     & All combined & Joint &   6.65 &    $>10$ &   7.68 &    $>10$ &   7.67 &    $>10$ \\
\bottomrule
\end{tabular}

    \caption{Comparison of the performance of the presupernova detection for the considered detectors using the counting analysis (CA), shape analysis (SA), and for the detector combinations using the joint shape analysis (Joint).  Detection threshold is  $\unit[5]{\sigma}$. Metrics are: $d(\unit[50]{\%})$ --- maximal supernova distance with $\unit[50]{\%}$ detection efficiency (top part); $t(\unit[50]{\%})$ --- amount of time to supernova, when the presupernova signal is detected with $\unit[50]{\%}$ efficiency from a \unit[200]{pc} progenitor distance (middle part); $z_{mean}(\unit[200]{pc})$ --- average presupernova detection significance for a \unit[200]{pc} progenitor distance, $\unit[1]{minute}$ before the core collapse (bottom part).}
    \label{tab:presn_results}
\end{table}

\newpage
\section{Summary}

We described an advanced method of real-time detection of a particular neutrino signal based on accounting for the signal's expected time profile. The proposed \emph{shape analysis} uses the log likelihood ratio as a metric for the hypothesis discrimination. This metric provides a pronounced enhancement of an experiment's sensitivity reach compared to the commonly used counting analysis. It also allows a fast and easy way to combine measurements of different detectors for a joint significance calculation.

The shape analysis method has been already implemented in the supernova neutrino trigger system of the NOvA experiment. The combination procedure applied to the two NOvA detectors increased the estimated sensitivity region of the experiment by $\sim\unit[1]{kpc}$.

Applied to the detection of the presupernova neutrino signal in KamLAND, Borexino and SK-Gd, the shape analysis provides an increase of the alert time before supernova by several hours.
As the results of the shape analysis depend on the chosen signal model, the highest significance is reached in the case of a model corresponding to the actual detected signal.
However, since (pre)supernova signal shapes share common features, even a wrong choice of the model gives a better sensitivity reach compared to the counting analysis.
Combination of different experiments and detection modes leads to a further sensitivity enhancement.

%Taking advantage of the expected distribution in time of neutrinos coming from either a presupernova or a supernova burst improves the reach of experiments to detect these important astrophysical events.  Comparing data from either different experiments or different channels in the same experiment further improves the distance at which such an event would be detected.  These techniques also allow better identification of the core-collapse time and earlier identification (and thus more advance warning) of the pre-supernova neutrinos.  Improvements are seen over a simple ``count the candidates in a rolling time window'' analysis for all model choices, as all available models produce time distributions substantially different than the ``constant'' distribution inherent in a counting analysis.

\acknowledgments{This work is supported by the Russian Science Foundation under grant 18-12-00271, and the US National Science Foundation under award 2013217. The calculations in this work were carried out using JINR cloud computing infrastructure~\cite{baranov_jinr_2016}.}
%\section{BACKUP}
%\newpage
%\todo[inline]{Backup}
%\input{sn_detection.tex}
%\newpage
%\input{sn_combine.tex}

\bibliography{sncombine.bib}
%\printbibliography
\end{document}